\documentclass[manuscript]{acmart}
\usepackage{subfigure}
\usepackage{balance}

\presetkeys%
    {todonotes}%
    {inline,backgroundcolor=yellow}{}
\AtBeginDocument{%
  \providecommand\BibTeX{{%
    \normalfont B\kern-0.5em{\scshape i\kern-0.25em b}\kern-0.8em\TeX}}}

\makeatletter
\renewcommand{\@thesubfigure}{(\alph{subfigure})\space}
\makeatother

\copyrightyear{2020}
\acmYear{2020}
\setcopyright{rightsretained}
\acmConference[RecSys '20]{Fourteenth ACM Conference on Recommender Systems}{September 22--26, 2020}{Virtual Event, Brazil}
\acmBooktitle{Fourteenth ACM Conference on Recommender Systems (RecSys '20), September 22--26, 2020, Virtual Event, Brazil}
\acmPrice{15.00}
\acmDOI{10.1145/3383313.3412213}
\acmISBN{978-1-4503-7583-2/20/09}

\begin{document}

%\title{Longitudinal Effects of Session-based Recommendations}
\title{Exploring Longitudinal Effects of Session-based Recommendations}

\author{Andres Ferraro}
\affiliation{%
  \institution{Music Technology Group - Universitat Pompeu Fabra}
  \city{Barcelona}
  \country{Spain}}
\email{andres.ferraro@upf.edu}

\author{Dietmar Jannach}
\affiliation{%
  \institution{AAU Klagenfurt}
  \city{Klagenfurt}
  \country{Austria}}
\email{dietmar.jannach@aau.at}

\author{Xavier Serra}
\affiliation{%
  \institution{Music Technology Group - Universitat Pompeu Fabra}
  \city{Barcelona}
  \country{Spain}}
\email{xavier.serra@upf.edu}

\renewcommand{\shortauthors}{Andres Ferraro, Dietmar Jannach, Xavier Serra}
\begin{abstract}
Session-based recommendation is a problem setting where the task of a recommender system is to make suitable item suggestions based only on a few observed user interactions in an ongoing session. The lack of long-term preference information about individual users in such settings usually results in a limited level of personalization, where a small set of popular items may be recommended to many users. This repeated exposure of such a subset of the items through the recommendations may in turn lead to a reinforcement effect over time, and to a system which is not able to help users discover new content anymore to the desirable extent.

In this work, we investigate such potential longitudinal effects of session-based recommendations in a simulation-based approach. Specifically, we analyze to what extent algorithms of different types may lead to concentration effects over time. Our experiments in the music domain reveal that all investigated algorithms---both neural and heuristic ones---may lead to lower item coverage and to a higher concentration on a subset of the items. Additional simulation experiments however also indicate that relatively simple re-ranking strategies, e.g., by avoiding too many repeated recommendations in the music domain, may help to deal with this problem.

\end{abstract}

%%
%% The code below is generated by the tool at http://dl.acm.org/ccs.cfm.
%% Please copy and paste the code instead of the example below.
%%

%\keywords{datasets, neural networks, gaze detection, text tagging}

%\begin{document}

%%
%% This command processes the author and affiliation and title
%% information and builds the first part of the formatted document.
\maketitle

\section{Introduction}
Recommender systems (RS) more and more determine which content we see in the online world. 
% DJX These systems can thereby have a major influence on the information that we receive and---at least to a certain extent---also on the decisions that we make. 
The exposure of selected items through recommendations aims at achieving certain desired effects like the discovery of %previously unknown content by users. 
new content. However, depending on the underlying algorithm, the repeated exposure of some items may also lead to undesired effects, like a too strong emphasis on already popular items. % \DJX in the recommendations.
Such ``blockbuster effects'', which may result in decreased sales diversity in e-commerce, were previously investigated for traditional recommendation approaches based on long-term preference profiles~\cite{Fleder2009Blockbuster,JannachLercheEtAl2015}, and they were also observed in field tests~\cite{LeeHosanagar2018}. Understanding such undesired effects is, however, important in many domains. %outside domains such as e-commerce. % DJX shortened
On news websites or on social media sites, for example, an over-emphasis on certain types of content may lead to a biased distribution of information, and to effects like filter bubbles~\cite{Pariser:2012}. 
% DJX moved this to another place.% TODO: Maybe add more (specific) references
%Solutions had been proposed to mitigate the popularity biases in the recommendations leading to a higher coverage by a small sacrifice of the accuracy \cite{abdollahpouri2017controlling,JannachLercheEtAl2015}. 
%\todo{Added the last sentence of previous paragraph}

% NOW: session-based recommendation.
The problem of such biases might be particularly pronounced in the context of \emph{session-based} recommendation scenarios, where we have to deal with anonymous or first-time users~\cite{hidasi2016session,QuadranaetalCSUR2018}. In such situations, % DJX: %which are very common and highly relevant in practice,
the recommendations by the system can be based only on the few observed interactions in the ongoing session. The level of personalization may therefore be low and lead to the effect that many of the provided recommendations consist mainly of generally popular items, eventually resulting in a reinforcement effect.
%Over time, the repeated recommendation of these items may then again result in a reinforcement effect.

This work is a first step to investigate such effects for the problem of session-based recommendation in a simulation-based approach.
% which was not done so far. %
%\todo{Added last part to previous sentence to make explicit that the contribution is to apply that approach to session-base recommendation}.
Starting from real-world data % TODO: Check if we have more than one at the end
of recorded user interactions from the music domain, we first generate session-based recommendations from a selected set of seed tracks with different algorithms. We then assume that a certain fraction of the recommendations are listened to by the users, and we correspondingly extend the dataset with these new interactions. This process is repeated in the simulation and from time to time we re-train the underlying models. After these retraining steps, we measure if the recommendations have changed on a system-wide level. In particular, we analyze if the recommendations have become more concentrated on a small subset of the items or not.
%\todo{added:} In this work, we simulate users interactions with the recommendations but we do not consider that users could also consume items that are not recommended by the algorithm. Even if this differs from a real world case, our goal is to focus on the effect that the algorithms would have if the users follows their recommendation. Therefore, we propose as future work incorporating in the simulations other interactions to reduce the effect of the algorithms.

Our experiments show that all investigated types of algorithms, both recent neural ones and heuristic-based ones, may lead to undesired concentration effects over time. Furthermore, we find that even though the prediction accuracy of some algorithms is often comparable, they may exhibit largely different concentration tendencies and, as a result, recommend very different sets of items to users in the end. This observation is particularly important from a practical point of view, since such differences might go unnoticed when an algorithmic comparison is solely based on accuracy measures. In practice, we are generally interested in a system that makes highly accurate predictions but does not exhibit undesired reinforcement tendencies. In an additional simulation experiment, we therefore investigated the effects when applying a re-ranking strategy to avoid too many repeated recommendations. The experiment provides indications that relatively simple strategies can help to counteract the undesired effects without a loss in accuracy.

% (The paper is organized as follows.) % TODO: DJ: I think we can skip this paragraph for a short paper.

\section{Previous Works}
\label{sec:previous-works}
%As one part of our research, we investigate how many of the available items are presented to users in their top-n lists by different algorithms. Herlocker et al.~\cite{Herlocker2004Evaluating} and others refer to this measurement as \emph{catalog coverage}. 
\emph{Catalog coverage} is a traditional quality factor in recommender systems and measures how many of the available items are presented to users in their top-n lists~\cite{Herlocker2004Evaluating}. In our work, we adopt the definition from~\cite{Herlocker2004Evaluating} and additionally measure how coverage develops over time.
%They propose to measure it by creating the union of the top-10 recommendations at a given point in time, and they emphasize that the metric should be combined with accuracy measures. We rely on their definition also in our work, but additionally measure how coverage develops over time.
In other works, % and in particular in the Information Systems literature,
catalog coverage is often referred to as \emph{aggregate diversity}, e.g., in~\cite{Adomavicius2012Aggregate}. One assumption is that higher aggregate diversity will ultimately lead to higher \emph{sales diversity}, as investigated for example in~\cite{LeeHosanagar2018}. In our work, we also assume that higher aggregate diversity (i.e., coverage) has positive effects.
Besides aggregate diversity, researchers also frequently investigate diversity at the individual level, see~\cite{Castells2015}. Aggregate and individual diversity are, however, not necessarily correlated~\cite{LeeHosanagar2018}. One can, for example, recommend the same set of highly diverse items to everyone, which does however not lead to high aggregate diversity~\cite{WangAdjustable2019}. 
%\todo{I think we can remove this text:} Questions of individual diversity are not in the focus of our work. % TODO: Could remove last sentence.

Various strategies were proposed to increase catalog coverage, typically based on re-ranking the top-n items returned by an accuracy-optimized algorithm~\cite{Adomavicius2012Aggregate,WangAdjustable2019}. Such strategies have the advantage that the resulting re-ranked lists remain highly accurate, i.e., only small compromises on accuracy have to be made. In our work, we investigate the effects of a comparably simple re-ranking strategy in a long-term perspective. % and additionally consider the long-term perspective. % DJX: % Our experiments also confirm that accuracy is not hurt by such a re-ranking.
Alternative model-based strategies for counteracting in particular popularity biases for traditional recommendation scenarios were proposed, e.g., in~\cite{abdollahpouri2017controlling,JannachLercheEtAl2015}.

% TODO: Check if there is actually a win as in \cite{JannachLercheEtAl2015b}

Using simulation-based techniques as a research method, e.g., in the form of agent-based modeling, has a long tradition in various fields, e.g., in managerial science~\cite{Wall2016}. Simulation-based research is however comparably rare in the field of RS. Recently, Zhang et al.~\cite{Gedas2020Longitudinal} used agent-based simulation to analyze longitudinal effects of recommender systems. Among other aspects, their simulations revealed a phenomenon called \emph{performance paradox}, where it turned out that a strong reliance by users on the recommendations may lead to suboptimal performance development over time. It was also found that RS can concentrate on a small set of items. Such concentration biases, also measured in terms of the Gini index, were previously explored in~\cite{JannachLercheEtAl2015}. Here, the authors as in~\cite{Gedas2020Longitudinal} observed in their simulation approach that some algorithms may lead to a concentration over time. However, some algorithms were also suited to increase catalog coverage and to decrease the Gini index over time. Our work continues the lines of research presented in~\cite{Gedas2020Longitudinal,JannachLercheEtAl2015,ferraro2019, chaney2018algorithmic}. However,
differently from previous works we consider session-based recommendation scenarios where no long-term preferences are available. %for users and where recommendations are correspondingly generated with the help of session-based recommendation algorithms.
% TODO: Could Discuss Prawesh2014MostPopular, if needed. They are, however, researching different problems like manipulation resistance or the penalty that an news article gets if it is, e.g., at position 11 but only 10 items are displayed.
% Would not discuss it here for space reasons.

\section{Methodology}
\label{sec:methodology}
Our general research methodology is based on simulation principles that were also adopted in~\cite{Gedas2020Longitudinal,JannachLercheEtAl2015}. As a starting point, % DJX for our simulations, 
we use datasets containing real user interactions. % DJX In our studies,
We focus on the music domain and use two public datasets %\todo{added:}
that include session information. One contains listening histories from the online music service \emph{last.fm} \cite{turrin201530music} called \emph{30Music}; the other one, called \emph{\#nowplaying}, was extracted from music-related Twitter messages in~\cite{Zangerle2014Nowplaying}. Regarding dataset characteristics, the \#nowplaying dataset comprises 1.2M listening events in 146k sessions for 61k items. The 30Music dataset is even larger, with 2.8M events 166k sessions for 446k items, i.e., sessions here are generally longer as well.
% TODO: Check if references are correct.
% TODO: one problem is that we don't know if these datasets contain ``organic'' listening behavior.
Note that both datasets contain user IDs and that long-term listening histories are available. Since we focus on session-based recommendation problems, we do not take these long-term models into account when recommending.

The main simulation procedure is as follows:
\begin{enumerate}
  \item For each session in the dataset, we select one track as a seed for a new listening session.\footnote{Alternative strategies with longer seed sessions are possible as well, but beyond the scope of this paper.}
  \item We generate playlists from the seed track using different session-based algorithms and measure the characteristics of the recommendations at a system-wide level.
  \item We assume that a certain fraction of the tracks in the playlist are listened to by the users and we add these simulated listening events to the dataset.
  \item We update the models at defined intervals and continue with Step (1).
\end{enumerate}

In our experiments, we used the following configuration:
\begin{itemize}
\item As seeds in \textbf{Step 1}, we randomly selected one of the tracks played in each session. We also made experiments with other seeds, e.g, the most frequently played track in a session. %track or the track that would receive the highest rating prediction by a matrix factorization algorithm.
    The obtained results were similar to the random seeds in terms of the general characteristics, which is why we omit them from this paper.
\item In \textbf{Step 2}, we created recommendations of list length 30. If all 30 tracks are listened, this roughly corresponds to a two hours music experience in case of pop songs. % TODO: find a good explanation for this choice. what is the average session length?
\item In \textbf{Step 3}, we assume that the users on average consume about one third of the recommendations, i.e., 10 tracks, based on the observations regarding adoption rates in the music domain in~\cite{KamehkhoschBonninJannach19,ferraro2019}. The selection of the tracks was done randomly. In general, modifying this parameter would result in slower or faster changes in the behavior of the recommender, but it would not impact the general characteristics of the emerging phenomena.
\item We retrain the models in \textbf{Step 4} after having generated artificial playlists 3 times. Assuming that models in real-world deployments are retrained over night, there would be 3 sessions per day before the models are updated.
\end{itemize}

Measurements are taken in Step 2 after each re-training step. To see how recommendations change over time for a given set of items, we repeatedly took the following measures for the tracks that were used in the first simulation round:
%\begin{enumerate*}[label=\textit{(\roman*)}]
\begin{itemize}
\item the \emph{Gini} index to assess the concentration of the recommendations on certain items. Higher values mean higher concentration~\cite{Gedas2020Longitudinal,JannachLercheEtAl2015};
  \item catalog \emph{coverage} in terms of the absolute number of different items appearing in the top-n lists;
  \item the average item \emph{popularity} in terms of the number of times a recommended track appears in the dataset;
  \item the information retrieval measures \emph{precision}, \emph{recall}, and F1 at list length 10 as accuracy measures.
\end{itemize}
%\end{enumerate*}

In our work, we seek to understand differences between algorithms in terms of their longitudinal effects. Depending on the application domain, the choice of the algorithm can then be based on these observations. We consider algorithms from different families in our simulations, as shown in Table~\ref{tab:algorithms}. The hyper-parameters of the algorithms were optimized for accuracy on the training data. %\todo{changed the following sentence to include that the parameters for the algorithms are there and the splits of the data as well}
To ensure reproducibility we share all code and data used in the experiments online, including the configuration for splitting the data and the parameters used for each algorithm.\footnote{\url{https://github.com/andrebola/session-rec-effect}}

\begin{table*}[h!]
  \centering
  \caption{Algorithms used in the Comparison}\label{tab:algorithms}
  \begin{tabular}{lp{12cm}}
  \toprule
  GRU4REC & The first widely-used neural approach to session-based recommendation, based on RNNS~\cite{hidasi2016session}.\\
  NARM & An attention-based neural method~\cite{Li2017narm}, often leading to competitive results~\cite{LudewigMauro2019}.\\
  SKNN & A nearest neighbor technique that shows competitive results in a number of domains~\cite{Ludewig2018}.\\
  CAGH & A simple yet often effective baseline proposed in~\cite{bonnin2014automated}, which recommends the greatest hits of artists that are similar to those appearing in the seed tracks. \\
%%  Spotify &  \textcolor{blue}{Maybe we leave this one out here.}
  \bottomrule
  \end{tabular}
\end{table*}

\section{Results}
\label{sec:results}
We report results for three experiments, discussed in Section \ref{subsec:results-initial-recommendations} to Section \ref{subsec:results-reranking}.

\subsection{Experiment 1: Analysis of Initial Recommendations}
\label{subsec:results-initial-recommendations}
%\textcolor{blue}{Report all for both datasets: precision, recall, gini, coverage, popularity (incl. relative change);
%see table 1 in other paper}
In the first measurement, we determine precision and recall for the different algorithms for the first round of recommendations, i.e., on the original data, and we additionally measure the Gini index, coverage, and the average popularity of the recommended items.

\begin{table*}[h!t]
\centering
  \caption{Results for first simulation round for the \#nowplaying dataset.}
  \label{tab:results-initial-nowplaying}
  \begin{tabular}{rccccccc}
    \toprule
Algorithm & F1 & Precision & Recall &  Gini & Popularity (abs.) & Popularity (rel.) & Coverage\\
\midrule
SKNN  & \textbf{0.1550} & 0.1482 &\textbf{ 0.1624}  & 0.4782 &57,7683 &39,8138 & \textbf{61,161}\\
NARM & 0.1481 & \textbf{0.1490} & 0.1472  & 0.5982 &65,7828 &48,0066 &59,578\\
GRU4REC & 0.1227 & 0.1175 & 0.1283  & \textbf{0.4169} &\textbf{ 22,7044} & \textbf{4,7920} & 61,119 \\
CAGH  & 0.0002& 0.0005 & 0.0001  & 0.9301 &171.7474 &153.6176 &24,718\\
%\rowfont{\color{gray}} Random & 0.0000 & 0.0002 &  0.0000 & 0.0667 & 17,9516 & -0.1179 & 61,220\\
  \bottomrule
\end{tabular}
\end{table*}

The results for the \#nowplaying dataset are shown in Table~\ref{tab:results-initial-nowplaying}. In terms of accuracy measures, we find that SKNN and NARM are working best in this experiment. %\footnote{The ranking of NARM and GRU4REC is different from the one reported in \cite{LudewigMauro2019}, because in our present experiment of a different measurement approach where we only create playlists from a single seed track for simulation purposes. This also explains the poor performance of CAGH.}.
GRU4REC works slightly worse in this setup, where we only use the first element of a session to create the playlist. CAGH, although often competitive in alternative setups, does not work in a satisfactory way in such a cold-start setup. Regarding the other metrics, we find that among the well-performing techniques, NARM has both the highest concentration bias and the strongest tendency to recommend popular items (see column ``Popularity (abs.)''). This is important because it shows that deep learning based techniques such as GRU4REC and NARM, despite comparable performance in accuracy, can recommend largely different items to users in their top-10 lists. %\footnote{We could run NARM only for five iterations before running out of memory. The tendency is however clear, and a simulation with 10\% subsamples of the datasets confirmed the trends observed on the full data.}
SKNN represents the middle ground here, but still leans quite strongly to recommend popular items. The column ``Popularity (rel.)'' shows the difference between the average popularity of the recommendations and the seed track. All algorithms recommend tracks that are more popular than the seed tracks, with GRU4REC being the best in terms of retaining the original popularity level. CAGH, by design, is worst here, as it only recommends greatest hits of artists. The differences in terms of \emph{coverage} among SKNN, NARM, and GRU4REC are low and all of them recommend almost all of the about 61k different tracks at least once. %Notably, the coverage of the random recommender is at about the same level. The coverage measures therefore seems to be not as informative as the Gini index, because even if one item only appears once in the thousands of generated playlists, it will increase this measure.
The results for the 30Music datasets shown in Table~\ref{tab:results-initial-30music} are similar in terms of the accuracy measures. On this dataset, however, SKNN is also favorable in terms of the beyond-accuracy measures.

%\todo{DJ: Discuss results of 30music once they are finished.}
\begin{table*}[h!]
\centering
  \caption{Results for the first simulation round for the 30Music dataset.}
  \label{tab:results-initial-30music}
  \begin{tabular}{rccccccc}
    \toprule
Algorithm & F1 & Precision & Recall &  Gini & Popularity (abs.) & Popularity (rel.) & Coverage\\
\midrule
SKNN  & \textbf{0.1988} & \textbf{0.1802}  &  0.2218  & \textbf{0.5629} &  \textbf{21,6084} & \textbf{15,5684} & \textbf{429,338} \\

NARM & 0.1955 &0.1697& \textbf{0.2306} & 0.7116 & 23.9657 & 17.9276 & 365,736  \\
GRU4REC & 0.1537 & 0.1318 &  0.1844 & 0.6547  & 24,0323  &  18,0633 & 397,470  \\
CAGH  & 0.0000 & 0.0000 & 0.0000  & 0.9340 & 83.2055 & 77.0883 & 141,257 \\ % \midrule
%\rowfont{\color{gray}} Random & 0.0000 & 0.0000  & 0.0000   &  0.1678 & 5.9608  & 0.0177  & 446,769\\
  \bottomrule
\end{tabular}
\end{table*}

\subsection{Experiment 2: Longitudinal Analysis of Concentration, Coverage, and Popularity Effects}
\label{subsec:results-longitudinal}
\begin{figure*}[h!]
\centering
\includegraphics[width=\textwidth]{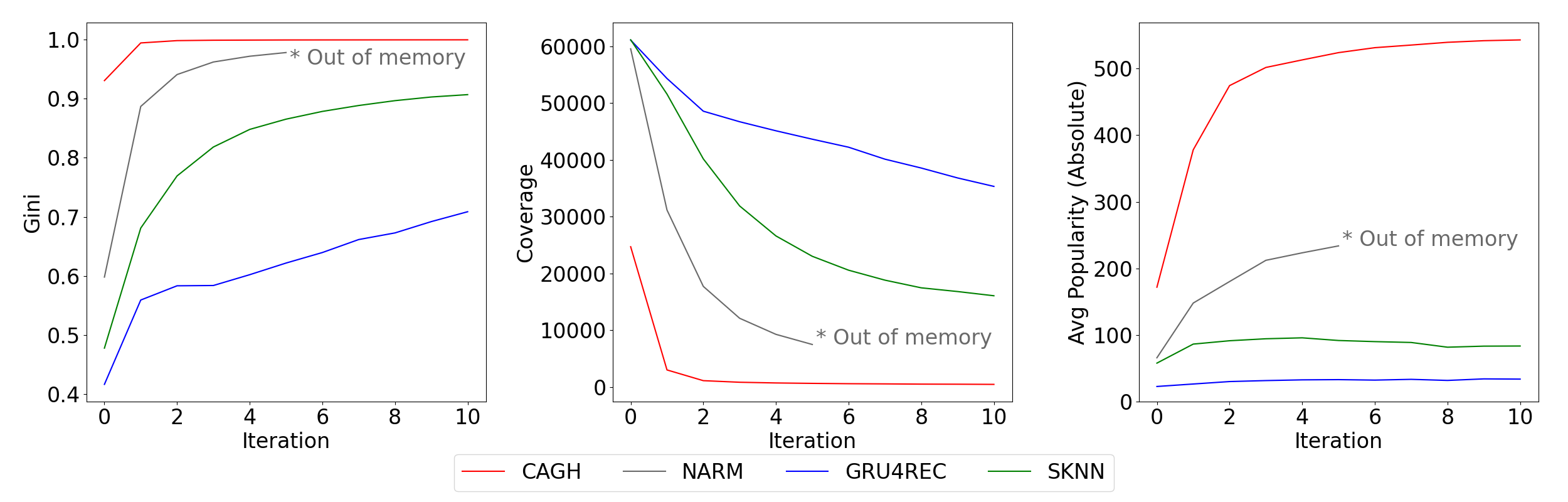}
\caption{Simulation Results for the \#nowplaying Dataset. NARM ran out of memory ($>$64 GB) after 5 iterations as we add more data to the training set. Additional simulations (not shown here) in which we created playlists for only 20\% of the data in each round confirmed the trends observed for the full datasets.}
\label{fig:fig-simulation-no-reranking-nowplaying}
\end{figure*}

% \textcolor{blue} {Show fig 3 for one of the datatests, only mention the other (including differences) to save space; report popularity developments only in text.}
In this experiment, we repeatedly generated playlists by randomly choosing seed tracks for each training session. In total, we made 30 simulation rounds. After each round we added the simulated interactions to the dataset and we re-trained the models in every third round, leading to the 10 iterations shown in Figure~\ref{fig:fig-simulation-no-reranking-nowplaying}.

Looking at the Gini index (Figure~\ref{fig:fig-simulation-no-reranking-nowplaying}), we observe that all algorithms in this comparison lead to an increased concentration effect over time. As expected from the results after the initial recommendations, the concentration increases most strongly when using NARM (excluding again CAGH), and it increases more slowly for GRU4REC. The development of the coverage metric follows this trend, as shown in Figure~\ref{fig:fig-simulation-no-reranking-nowplaying}, i.e., the coverage of all algorithms goes down steadily during the simulation, with the strongest effect observed for NARM and the weakest for GRU4REC. Interestingly, the average popularity level remains mostly constant for SKNN and GRU4REC, with GRU4REC generally having a lower popularity bias than SKNN. For NARM, the popularity bias however increases over time.
%\todo{AF: Re-do the analysis for NARM here for a subsample of the data. And report this problem of running out of memory.}

%We run all the experiments with 64GB of memory but for the case of NARM we could not retrain after the fifth round because it requires much more memory than the others algorithms. Therefore, we sub-sample the sessions to only a 20\%.

\begin{figure*}[h!]
\centering
\includegraphics[width=\textwidth]{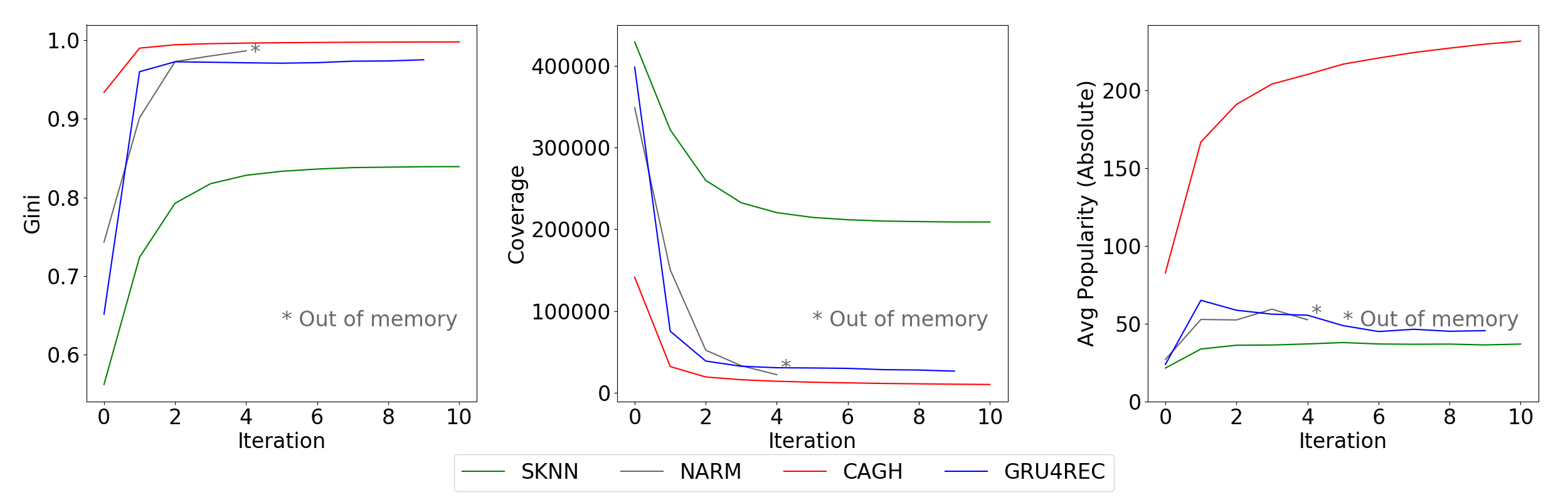}
\caption{Simulation Results for the 30Music Dataset. NARM again ran out of memory after a few iterations.}
\label{fig:fig-simulation-no-reranking-30music}
\end{figure*}

The simulation results for 30Music are shown in Figure~\ref{fig:fig-simulation-no-reranking-30music}. The general observations are similar to those obtained for the \#nowplaying dataset, i.e., all algorithms lead to a concentration over time and to a reduced coverage. As expected from the results of the initial recommendations (Table~\ref{tab:results-initial-nowplaying} and Table~\ref{tab:results-initial-30music}), we can however see that SKNN behaves favorably on this dataset in terms of concentration and coverage.

%there are some differences across the datasets with respect to individual algorithms. In particular the concentration effect of GRU4REC is much higher on this dataset, which indicates that certain developments over time depend on the characteristics of the underlying datasets.
%\todo{DJ: complete the discussion once the results for NARM are here. Also report some dataset statistics.}

\subsection{Experiment 3: Longitudinal Effects of Using Reranking as a Countermeasure}
\label{subsec:results-reranking}
%\textcolor{blue} { Fig 4 for one of the datasets, using maybe the user-independent reranking strategy; show effects on accuracy. Maybe plot accuracy on same graph (right hand size axis labeling) on all graphs if possible.}

In a final set of experiments our goal was to investigate the effectiveness of applying reranking strategies to avoid concentration effects as was done, e.g.,~\cite{Adomavicius2012Aggregate}. We analyzed the effects of two basic heuristics:
\begin{itemize}
\item In \emph{Reranking Strategy 1}, our goal was to avoid recommending tracks too often that were recommended frequently in previous rounds to all users. Technically, in each round we count the number of times an item $i$ was recommended, denoted as \emph{previous\_recs(i)}. %In the following round, for each item recommended in a session ($s$), we re-order the recommendations as described in \eqref{eq:1}.
    In the following round, we penalize frequently recommended tracks by moving them back in the recommendation lists. The penalty $p$ in terms of the number positions to move the item $i$ back is computed heuristically as $p = 10*log(\text{previous\_recs}(i))$.
%\begin{align}newposition(i,s)  = originalposition(i,s) + 10*log(\text{precs(i)})\label{eq:1}
%\end{align}
\item \emph{Reranking Strategy 2} is personalized, and it tries to avoid recommendations that an individual user has consumed previously in the same session. %Note that while our focus is on session-based recommendation, where long-term information is not generally available for all users, this experiment gives us some insights on the possible benefits of personalization.
Technically, for each user $u$, we count the number of times it has consumed an item $i$ in the session, denoted as \emph{previous\_consumptions(i,u)}. The penalty $p$ for user $u$ and item $i$ is consumed as $p= 10 \times \text{previous\_consumptions}(i,u)$. For example, if a user has listened to a track two times before, the track will be moved back 20 positions in the new recommendation list.
    %we re-order the recommendations following \eqref{eq:2}. E.g. if the user listened to an item two times it will be moved 20 positions lower in the recommendation.
%\begin{align}newposition(i,s,u)  = originalposition(i,s) + 10*\text{precs(i,u)}\label{eq:2}\end{align}
\end{itemize}

%Figure \ref{fig:fig-simulation-reranking-nowplaying}.

\begin{figure*}[h!]
\centering
\subfigure[Reranking based on Recommendation Frequency]{
%\rule{.5\textwidth}
  \centering
  \includegraphics[width=.50\linewidth]{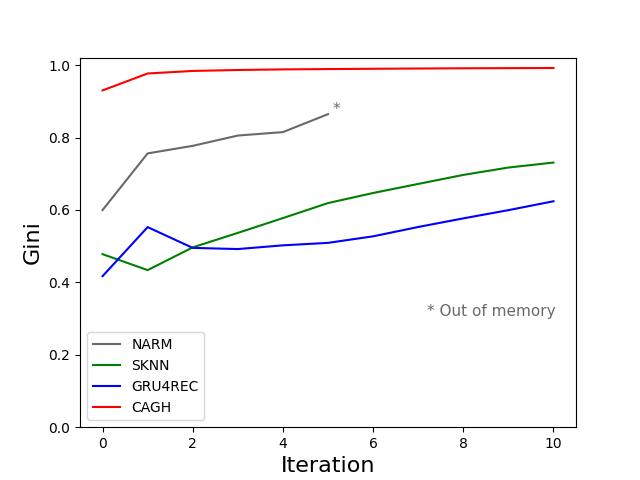}
  \label{fig:nowplaying-simulation-gini-rerank}
}~
\subfigure[Individualized Reranking based on Consumption Frequency]{
%\rule{.5\textwidth}
  \centering
  \includegraphics[width=.50\linewidth]{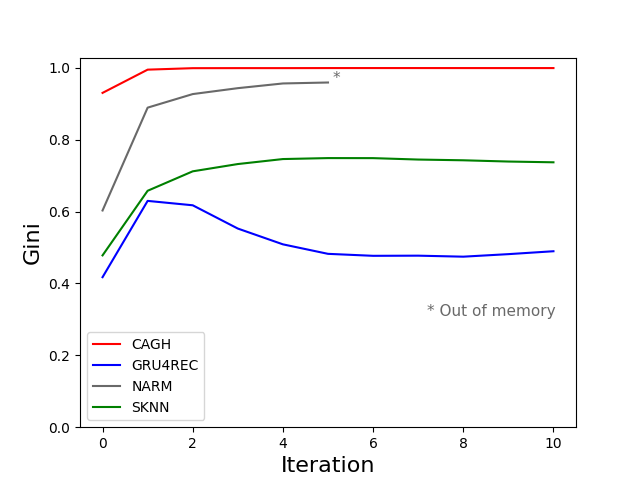}
  \label{fig-simulation-nowplaying-reranking-user}
}

\caption{Simulation Results (Reranking) for the \#nowplaying Dataset.}
\label{fig:fig-simulation-reranking-nowplaying}
\end{figure*}

The results show that both re-ranking strategies are effective, but in slightly different ways. When avoiding to repeatedly recommend the same tracks to everyone (\emph{Reranking Strategy 1}), we can observe that the increase in concentration can be slowed down for all algorithms except again for CAGH (compare Figure~\ref{fig:nowplaying-simulation-gini-rerank}). For the case of individualized reranking \emph{Reranking Strategy 2}, the increase of the concentration bias can be stopped at a certain level for SKNN and GRU4REC. For NARM, however, which exhibited a strong concentration bias already at the beginning, this personalized reranking does not seem to be very effective. This phenomenon can be attributed to the limited level of personalization of NARM, as shown in Table~\ref{tab:results-initial-nowplaying}. Overall, however, the results also indicate that already simple reranking strategies can be effective countermeasures to avoid undesired concentration effects.
%\todo{AF: Do we have accuracy results for the re-ranking? How much do we lose in precision and recall after re-ranking? Please add some numbers here (one or two sentences only).}
%\todo{Added paragraph explaining this:}
%If we average the precision and recall across all rounds in our simulation,

Furthermore, the reranking strategies do also not lead to a loss in accuracy.
% which often come with a limited loss of accuracy \cite{Adomavicius2012Aggregate} or sometimes even to a slight increase \cite{JannachLercheEtAl2015b},
Looking a the precision and recall values obtained in our simulation experiment, we see that the accuracy of GRU4REC for both reranking strategies remains almost constant; for SKNN, the performance is even slightly increased, as observed previously for the music domain in~\cite{JannachLercheEtAl2015b}. These results are consistent for both datasets. %This shows that for SKNN we get the highest accuracy and at the same time we can improve coverage.
%\todo{Probably we don't want to include this, but I added in case we want to add more details:}
%Note that the accuracy values vary on each iteration.
Specifically, if we average precision and recall over all iterations for GRU4REC without reranking on the \#nowplaying dataset, both precision and recall are at about 0.11. Applying either reranking strategy only leads to changes at the third place after the decimal. For SKNN, precision and recall even go slightly up with both strategies from about 0.13 up to 0.16.

%we obtain 0.1161 for precision and 0.1195 for recall. With \emph{Reranking Strategy 1}, precision is 0.1107 and recall is 0.1114. With \emph{Reranking Strategy 2} the precision is 0.1135 and recall is 0.1158.
%If we compare the same results for SKNN we see an increase in the accuracy. Precision without reraking is 0.1348 and recall is 0.1585. With \emph{Reranking Strategy 1} the precision is 0.1590 and recall is 0.1599. With \emph{Reranking Strategy 2} the precision is 0.1507 and recall is 0.1616.

Finally, additional measurements, which we omit for space reasons here, show that coverage also ceases to go down for SKNN and GRU4REC when a personalized reranking strategy is applied, and that the popularity bias continues to remain stable. The same measurements were furthermore made for the 30Music dataset (except for NARM, again due to high computational costs). The results are again generally in line with what was observed for the \#nowplaying dataset.
%\todo{AF: Do we have results for the other dataset as well?}\todo{Yes, they are in the other latex document in a table. I think is not necessary to include them since they are similar to nowplaying}

\section{Conclusion}
In this work we analyzed through a simulation-based approach to what extent modern approaches to session-based recommendation exhibit certain potentially undesired biases, e.g., recommending the same set of items to everyone, and if these biases are reinforced over time. Our analysis in the music domain shows that, unlike to the findings in~\cite{JannachLercheEtAl2015}, all examined approaches can have such tendencies, although to a different extent. For practitioners, our work implies that such effects should be considered in the choice of algorithms, in particular because accuracy measures alone do not reveal such important differences between algorithms. Furthermore, our first experiments with personalized and non-personalized reranking strategies showed that already simple heuristics can be helpful countermeasures. On a methodological level, we see our work as another step to move beyond today's ``single-snapshot'' analysis of recommendation algorithms, which does not allow us to investigate or simulate longitudinal effects of such systems.

Our work so far is limited to one particular domain and a specific set of assumptions used in the simulation, e.g., that the observed interactions all come from the recommendations, which leads to an amplification of the observed effects.
Our future works include the investigation of other domains like e-commerce and the consideration of scenarios where long-term preference information about the users can be leveraged to diversify and de-bias the recommendations.  %DJX %\todo{added:}For different domains, other penalties for the re-ranking can be applied to evaluate the generalization of the proposed solution.

%%
%% The acknowledgments section is defined using the "acks" environment
%% (and NOT an unnumbered section). This ensures the proper
%% identification of the section in the article metadata, and the
%% consistent spelling of the heading.
\begin{acks}
This research has been partially supported by Kakao Corp.
\end{acks}

%% The next two lines define the bibliography style to be used, and
%% the bibliography file.
\balance
\bibliographystyle{ACM-Reference-Format}
\bibliography{acmart}

%%
%% If your work has an appendix, this is the place to put it.

\end{document}